# A Visual Query System for Scholar Networks

Hongze Li [1]

[1] Lhongze@hotmail.com

**Abstract:** Large scholar networks is quite popular in the academic domain, like Aminer. It offers to display the academic social network, including profile search, expert finding, conference analysis, course search, sub-graph search, topic browser, academic ranks and user management. Usually the search results are listed as items, while the relations among them are hidden to the users. Visualization is a feasible way to help users explore the hidden relations and discover more useful information. This article aim to visualize the search results in Aminer in a more user-friendly way and help them better utilize the tool. We provided three different designs to visualize the results and tested them in user study. The empirical results of our research show that the designed graphs help users better understand the area they intend to know and make their search more effective.

## 1. Introduction

### 1.1. Research Background introduction

With the acceleration of people's living rhythm and the increase of social complexity, more and more data are appearing in our sights. These data come from life as well as life. People also pay more and more attention to the importance of research data. Necessity, but the form of the data looks complicated and cumbersome, can not properly grasp the data features and data description of the problem, no doubt, through visualization [1], will be displayed directly to our data list into a visual image of people can feel, symbols, colors, textures, etc., enhance the efficiency of identifying large amounts of data. It is a good way to visualize data in front of our eyes and it is also easy to capture information that the data wants to convey to us. Information visualization is closely related to human perception and human behavior, especially human interaction[2].

The development of the Internet makes search engines become the most important means for users to find information, and accurate and rapid is the user's most important requirement for search engines. The original search engine and retrieval system are only displayed to the user in the form of a list according to the results required by the user. The lack of interaction and lack of intuition can not find the relationship between the search results, let alone the search process Interesting.

### 1.2. Research content introduction

Today's visualization has penetrated into all fields, and large scholar networks are very popular in the academic field. It provides a display of academic social networks[3] including profile search, expert discovery, meeting analysis, course search, topic browser, academic ranking and user management.

Based on the search engine[4], we can more easily obtain the information of the scholars. The query results are typically listed as items, to describe the expertise of each scholar and their academic performance. However, the scholar networks typically consist of millions of scholars with doses expertise. Organizing the results is highly demanding. Ranking is proposed to help for this kind of organization. It provides quite promising sequence to assist users to check the scholars. However, when querying the certain scholars, a number of criteria involved that makes it difficult to rank the scholar in a one dimensional line. Also, users wish to gain more insight in terms of the relationship among the scholars - to help for classification, search for collaboration, and even scientific study. The ranking cannot achieve this to capture the interrelationships behind them.

To better demonstrate this problem, we propose to utilize the visualization to help users for the multi-criteria search. Visualization is a feasible way to help users explore hidden relationships and discover more useful information.

The scholar's information retrieval system is the application of visualization in the field of search[5]. The Echarts and d3.js visualization libraries used are designed as the basis for the visualization of the page. Three visualization forms are designed, and the colors and lines are distinguished. The search results are more intuitive and interactive, with more fun and readability, and the functions are expanded and enhanced. The user can use the retrieval system to conveniently retrieve the required keyword information, and help to clarify the relationship between the sub-concept and super-concept words between the keywords. At the same time, the expansion of the new word and the double click event under the event can be used to retrieve keywords[6]. Greatly improved the search efficiency. The visual search system is now generally recognized by everyone. More and more search websites in the world have begun to try to use the visual interface[7], so this is the trend of the times.

Our aim is to obtain keywords-related information by searching for keywords and selecting keywords, such as the relevant scholars in the field and the historical development of the field. Therefore, our visual design is better able to realize understanding of the field and related fields before we want to understand a certain field, form an overall relationship network, conduct more in-depth study of key areas, and our system through visualization. Presentation and convenient interaction achieve this goal.

Our paper is organized as follows. Section 2 presents the related work. Section 3 gives an overview of the system. Section 5 and 6 describe the core part of the system - graph based query system and show the interaction of the visual query system respectively. Section 7 provides different user cases. Section 8 ends with conclusions.

## 2. Related Work

Most of scholars' search system's Knowledge Graph is a check box form to select keywords. Not only is the relationship between keywords unique, but also the sub-concept and super-concept of some keywords are hidden[8]. The user searches for terms related to super-concept. For example, the AI's super-concept is computer science, and the sub-concept is the search keyword's subordinate concept. That is, the narrower, related term, for example, the AI's sub-concept has automatic Driving, etc. When we expand the word when we search for an abbreviation, for example, the acronym for Artificial Intelligence is AI, then its expansion is Ab Itinio.

When searching, you need to search the box again to perform an input search. These steps are not only complicated but also not clear to the user.Due to the continuous increase and improvement of data, the results displayed by search have become more and more complicated.The original sorting method can not display the results intuitively, and the selection method is not easy to understand. It is just the result of textual display, and hides the relationship between keywords. It is very impractical. We want to design a visual system to show more results. This can be applied not only to scholars' search but also to more search systems. In different use cases, a visual display system can be dynamically designed based on data[9]. There are many methods of visualization that can be used.The purpose is to make the system visual display, while adding some interactive features to provide human-computer interaction experience and enhance the search function.

### 2.1. Visual display part design

Considering the layout of the page and the usable range of the page, select the most suitable visualization in the Echarts and d3.js libraries, and finally select the interesting and relationship-connected force-oriented diagram in d3.js. Display extended word information[10], and at the same time find a tree diagram in Echarts that has an up-and-down relationship and has nodes to display the super-concept and sub-concept words, and imagine that it can be color-differentiated[11]. Set two regions in the display area. If the user searches for an abbreviation, the extended word module will automatically display the extended word. If the user searches for a complete word, the extended word module shows the background image. The word search result display module displays the results of the sub-concept and super-concept of the keyword.

### 2.2. Interactive Design

Interaction is an important part of our visual query system[12]. It not only makes searching easier, but searching is more interesting. Most web pages focus on perfect functionality, but rarely consider the user's experience. With the popularity of visualization[13], web pages should also be more intuitive and easier to use[14]. Our system takes into account all aspects of user requirements.

First we can use the click tree node to get the user's desired sub-concept word, but if this node has already acquired the child node, then our click will be the expansion and contraction of the node. To improve usability and differentiation, we designed a number of changes in colors and fills[15]. Click on the interaction we can only receive the sub-concept word information of this keyword, but many users want to find out the word and the expansion of the word, so we have designed a double-click event. A double-click event will search for upstream and downstream word information for this node. Second, we added two MORE nodes to dynamically add the number of super-concept and sub-concept words, because to meet the needs of different users, this node is completed by clicking.

## 3. System Overview

### 3.1. The original search interface

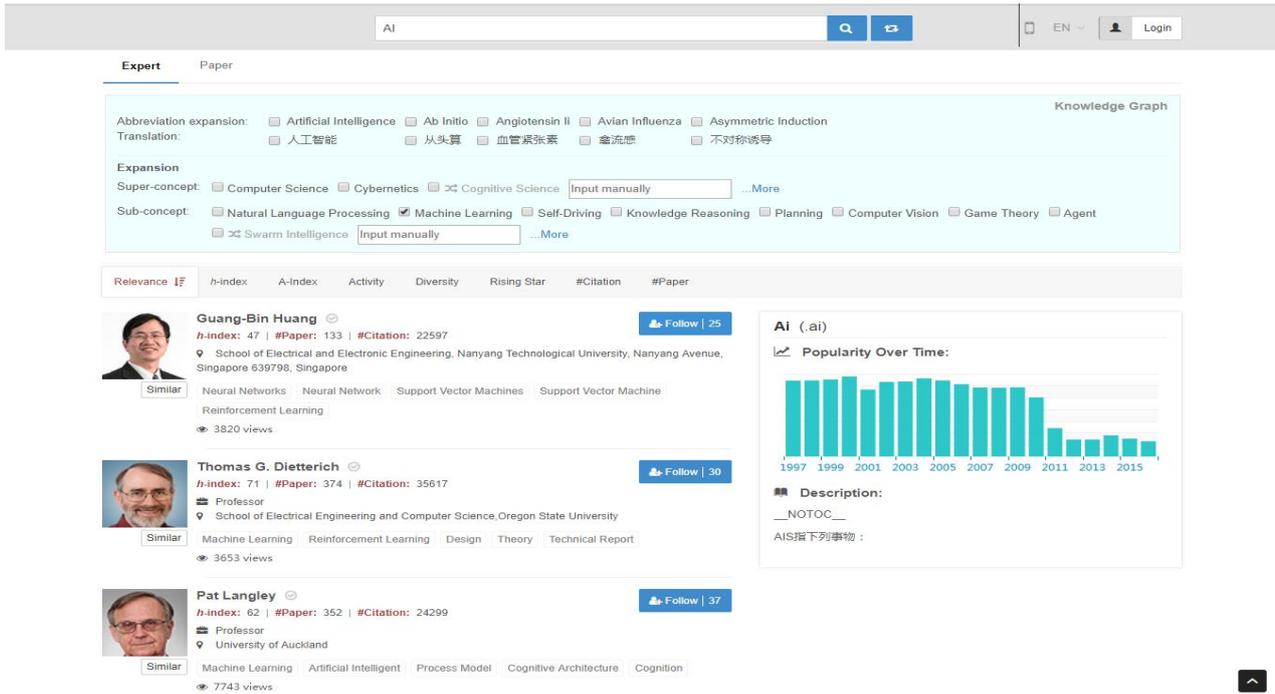

Figure 1. The graph is the original search interface.

The original search interface is a list of pages, and select the keyword by Check box[16]. More and more people see that check boxes are boring and feel complicated. When we want to search a key word, we always have a headache and dazzled. While the relations among them are hidden to the users[17]. So when users want to search a key word's Super-concept and Sub-concept they must write the word into the search bar. This method is very complicated. So we want to improve it. In this example, we have selected machine learning, and then the system shows relevant scholars in the scholar network[18].

To improve the system, we need to understand the operating mode and internal structure of the original system. Therefore, we explored that the original system works by searching keywords and classifying keywords, including sub-concept, super-concept, and extensions. These results are presented in sub-cases, provided to the user for selection, and then relevant information can be obtained through these keywords, such as displaying scholars related to this. Its working method is shown in the figure below.

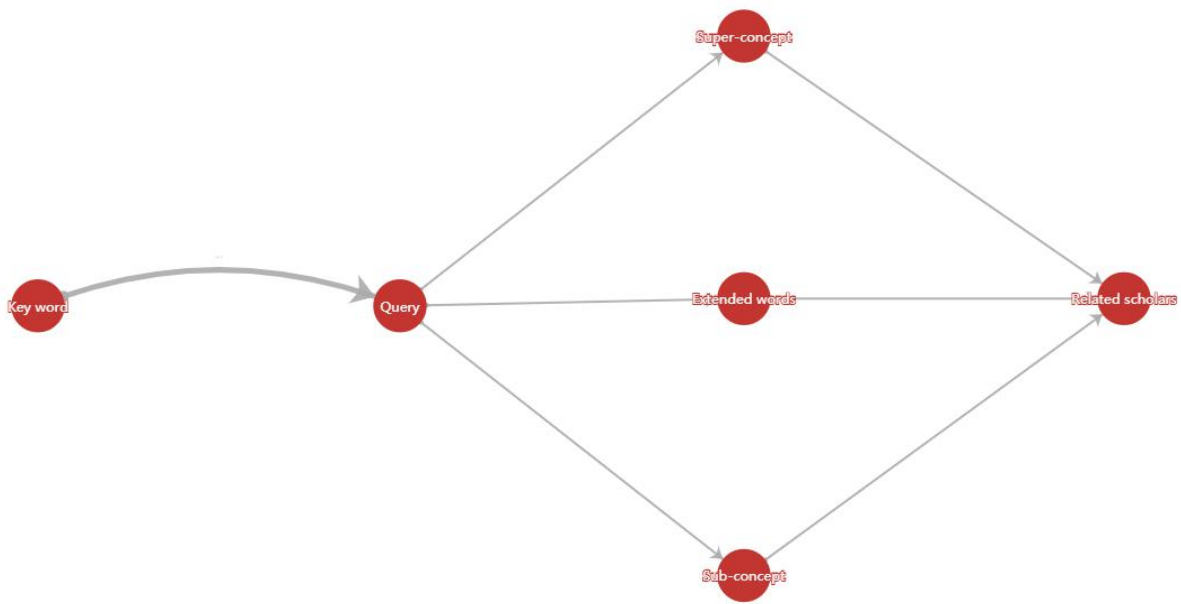

Figure 2. The system internal process of the scholar query system.

3.2. Visual Query System

We designed a visual query system to search the key words. We aim to visualize the search results in Aminer in a more user-friendly way and help them better utilize the tool[19]. Our system consists of mainly four parts - Query Bar, Keyword Expand Panel, Query Graph Explorer and Query List Display. The Query Bar on the top mainly contains the bar that allows users to type their keywords for search. The Keyword Expand Panel on the left up corner shows the options of the keywords. The Query Graph Explorer on the right up corner demonstrates the process of the query. The Query List Display on the bottom shows the corresponding scholars and their properties.

This system show that the designed graphs help users better understand the area they intend to know and make their search more effective. And the way people interact with the page is more interesting and can make it easier for people to understand the relationship between keywords. Utilize our Visual Query System to display the scholars who need to understand more information.

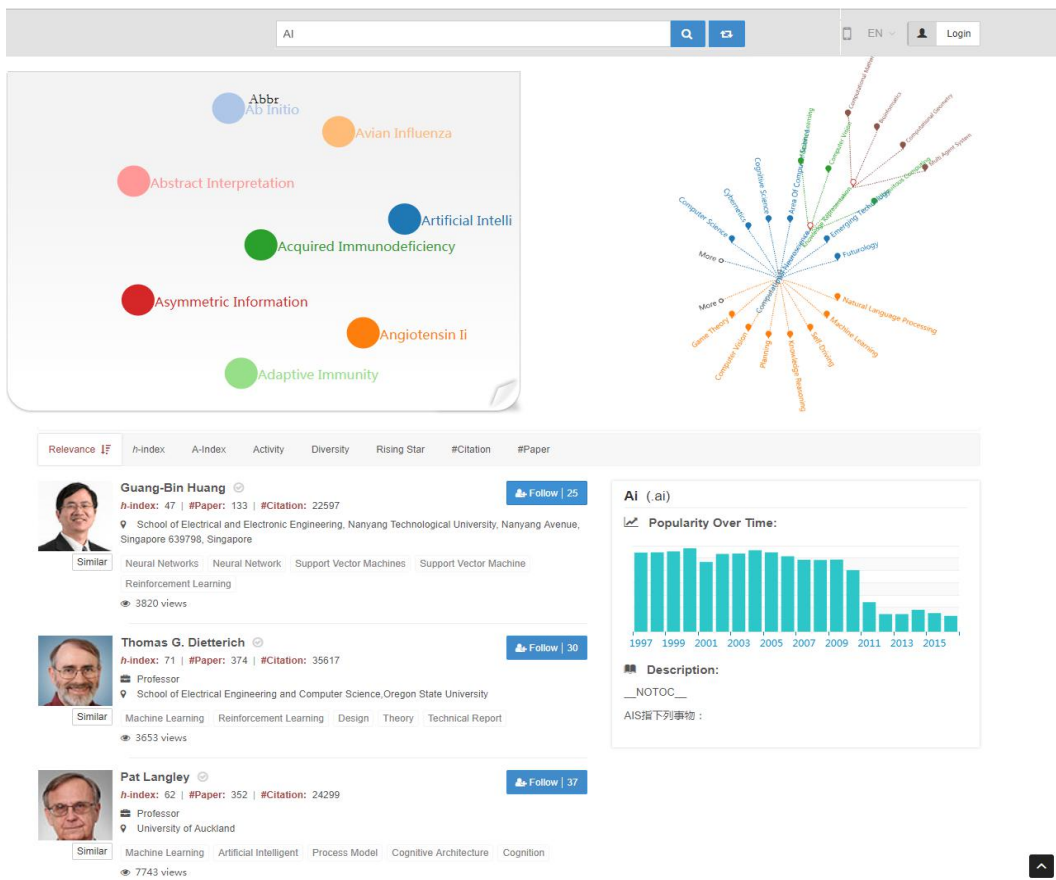

Figure 3. The visual query system of scholar.

3.3. Keyword Expand Panel

It is quite common that some keywords are abbreviation of the original words. For example, "AI" represents the "Artificial Intelligence" in the computer science domain, but it also means "Asymmetric Information" some fusion occurs. To better support the query, we provide the "Keyword Expand Query". User could first types in the abbreviation of the keywords, then some original terms are proposed to the user. He/she can pick the extract keyword.

Extended words allow us to extend a basis for search. So we designed a module to display extended words. When we search the abbr like AI, our system can find and show extended words through force map. The significance of the existence of extended words is to give users more tips, so that they can associate more relevant keywords, provide users with more ideas to get more knowledge.

We also add some events in the extended words to facilitate users to obtain relevant information conveniently. For example, by clicking on the circles of various colors to realize the search function, more information about the clicked keywords is displayed to the user.

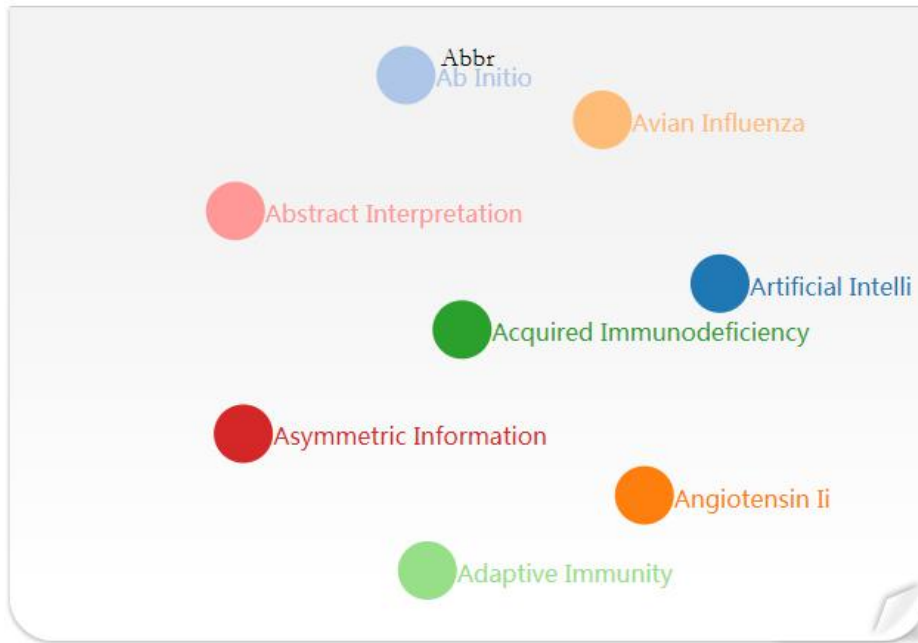

**Figure 4. The extended word display module in visual query system.**

## 3. Query Graph Explorer

As we production three different methods to show the key words of the scholar networks. Since different users have different preferences, we could choose different methods to demonstrate results.

4.1. Radial tree

We mainly studied Radial tree.Because this tree map is very beautiful and it can easily to extended more features. We can saw the super-concept are displayed in blue in the upper part and the sub-concept in orange In the lower part.This layout is more intuitive and easy for users to accept and understand[21]. Then we designed two nodes named More to add super-concept and sub-concept.

The advantages of radial tree are the following points. Fast display. The data display is multiple layers. Under the visual analysis, the data is divided into multiple layers and the layers are extended to each layer. Each layer represents a keyword layer. The circular layout extends outwardly to more intuitive display of relationship information, which looks intuitive and enhances the user's understanding of keywords. Many of the key words covered in the graph are related of multiple elements. One element will affect many other elements,If we do not take a visualization, you will not be able to see the whole picture and you will not be able to have a real Hierarchical relationship discussion.

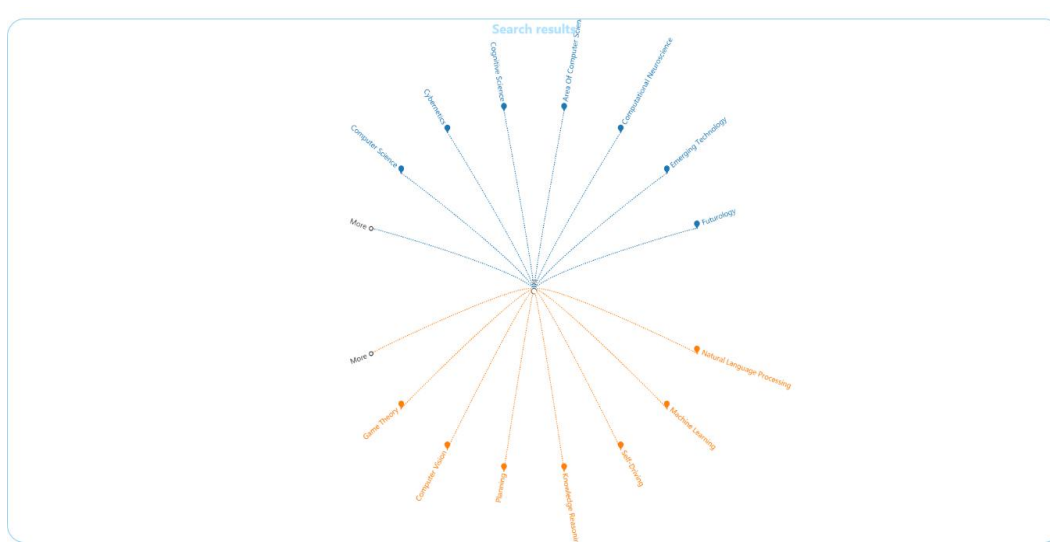

Figure 5. One of the display methods of the visual query system: Radial tree.

## 4.2. Horizon tree

The second method is a horizon tree, this kind of tree map is easier to distinguish super-concept and sub-concept. The first requested words may not be what the user is looking for, so click the More node until find the result.

There are also many advantages of the horizontal tree diagram. For example, this kind of display format saves the space of the page and facilitates the layout of other modules in the system. This method also shows the relationship between keywords in a more intuitive manner, facilitating the user to understand the keyword information. However, the level of this method is not good. When there are many open words, the page is complicated and not easy to watch.

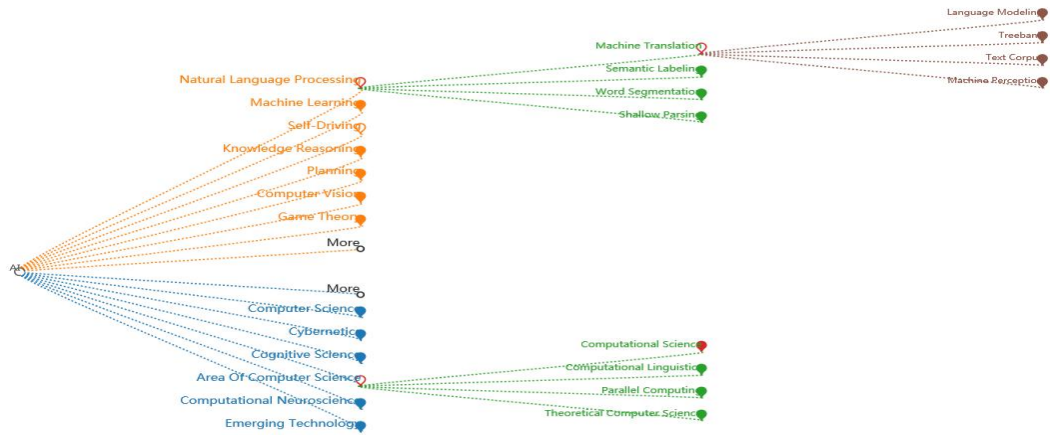

Figure 6. One of the display methods of the visual query system: Horizon tree.

## 4.3. Force map

We have designed a force map, we can put colorful balls in this module we want to put according to the needs of the system and the needs of the users. We plan to add a mouse click event to each circle and click to show the related keywords. More information and increase the user experience.

Mandatory maps have many advantages. The most important is that it is a more interesting layout. These nodes have a richer color. We can drag it. The interaction is very strong, but its display information capability is not the first two kinds of intuitive. Keywords The relationship between the two is also not strong enough, so there is still room for improvement.

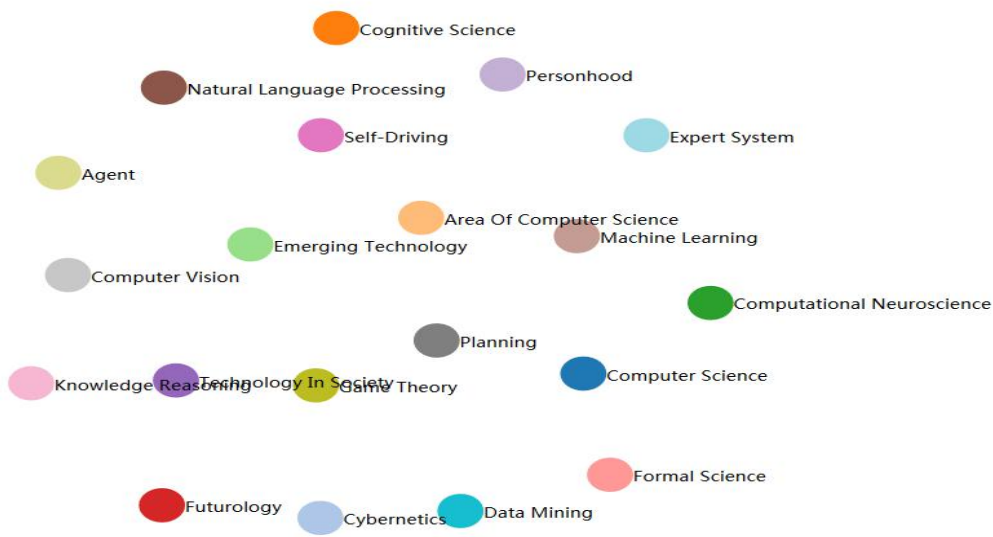

Figure 7. One of the display methods of the visual query system: Force map.

## 5. Visual Query Interaction

Interaction is the important part in our visual query system. Not only does it make it easier to search but it can be more interesting to search[22]. Most web pages focus on whether the functionality is perfect, but rarely take into account the user's experience. With the popularity of visualization, web pages should also be more intuitive to display and more convenient to use. Our system takes into account all aspects of the user's needs. Form a multi-level web page[23].

5.1. Expand the Graph

We could use the click tree nodes to obtain the sub-concept that the user wants to. For example, emerging technology is the super-concept for AI, we could click the Emerging technology nodes to receive this word's sub-concept. Like the green nodes and green lines in this graph. So we could use the same method to receive the sub-concept of green nodes. Like the brown nodes and brown lines in this graph. This figure is all the nodes can click interaction. But if this node already has children, then our click will be the expansion and collapse of the node. In order to improve usability, we designed many color and fill changes. When the node our click is not has sub-concept the node keeps its color unchanged and becomes an empty node, like the node self -Driving. When the node our click has sub-concept the node becomes a red empty node, like the node Emerging technology and Machine learning. Considering that there are too many nodes in the tree graph will become more complex, we can click to get the descendant of the child nodes, then shrink the node and the node become red fill node, like the node Natural Language Processing.

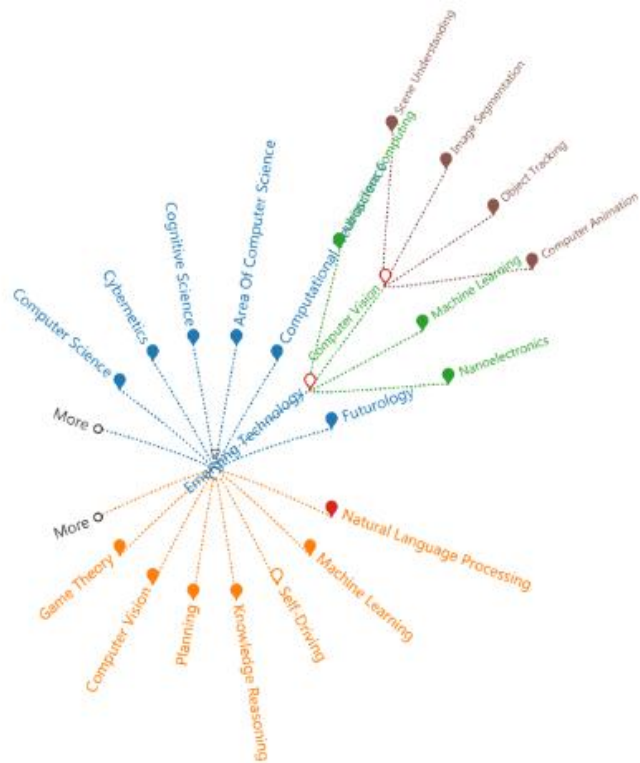

Figure 8. The method for expanding keywords in our visualization system: Click interaction.

5.2. Get more information on keywords

Click can only receive sub-concepts of this key word, but many people want to find out this word's super-concept and extended words. So we designed double click event. As we double-click the super-concept of AI, like node emerging technology. At this point, the node emerging technology is passed to the center of the tree map, and the emerging technology super-concept and sub-concept are expanded around. Also we can click and double-click interaction in this figure.

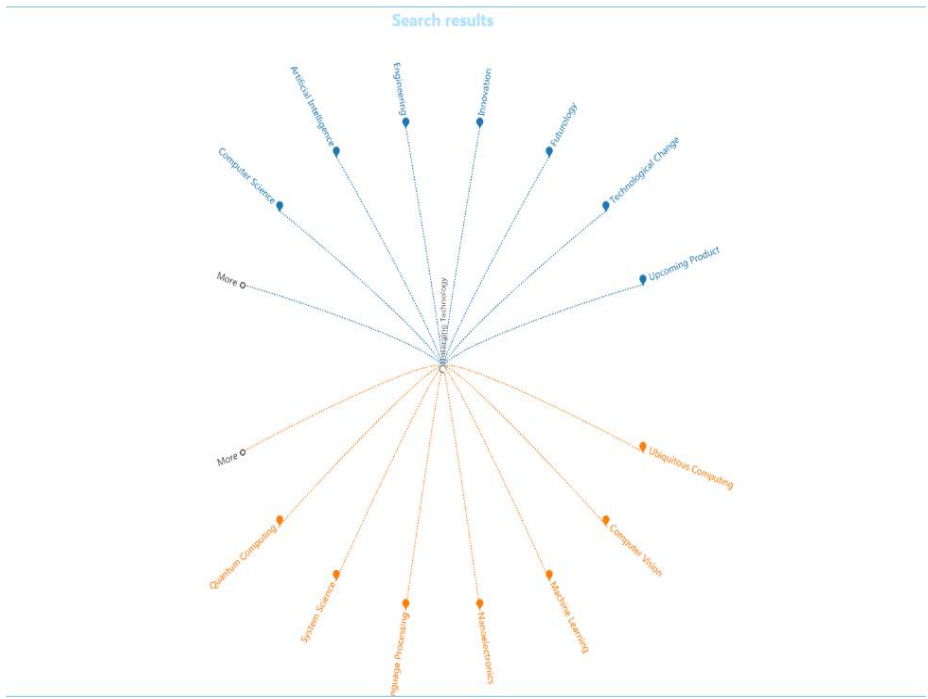

Figure 9. The method for expanding keywords in our visualization system: Double-click interaction.

### 5.3. Case Study

#### 5.3.1. Data Mining

We can not only search for abbreviations to get keyword information, but also we can get information on complete keywords, like the keyword Data mining. But in this we don't receive expand words. Also we can click and double-click interaction in this figure.

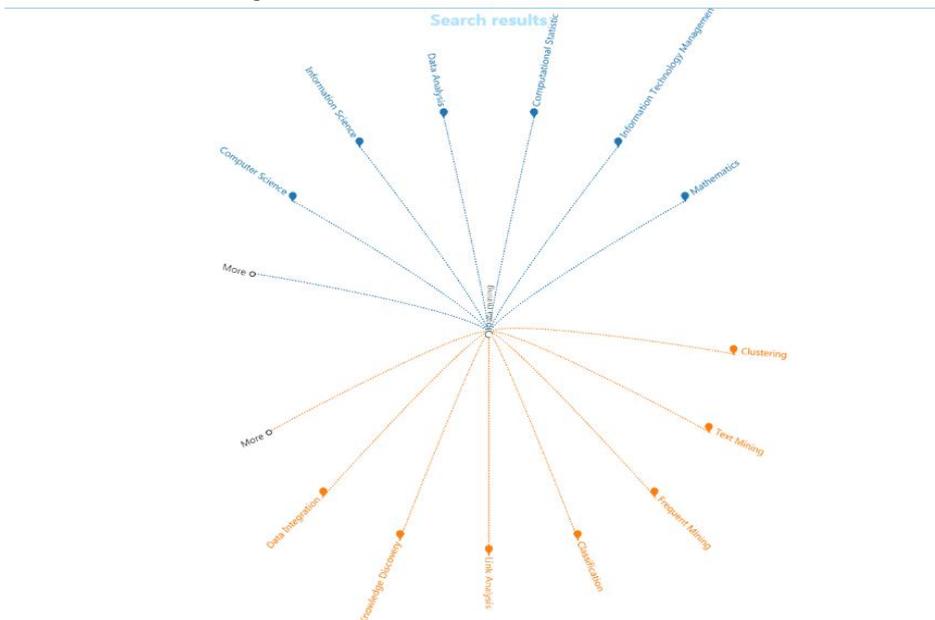

Figure 10. Search the full name of the method to use the visualization system.

#### 5.3.2. Data Integration

The biggest advantage of our system is that it can accommodate more keyword information and meet the needs of all users. After a lot of actual testing, we learned about various special situations because we have considered various situations to design the system. An error occurred on the system and the system stopped working.

Some keywords have no sub-concept, so when we search for such words, only the super-concept words are shown, like Data integration.So we can saw the super-concept of the Data integration and we could click the nodes to expand this node's sub-concept. This situation give user more select to Interact with the computer.

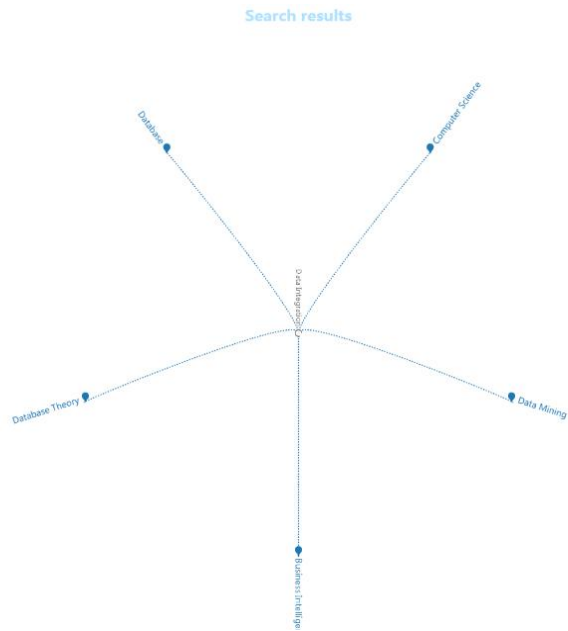

Figure 11. Some keywords only the super-concept words are shown, like Data integration.

### 5.3.3. Knowledge reasoning

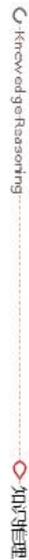

Figure 12. Some keywords only the translate words are shown, like Knowledge reasoning.

The translation function can be an independent system. Although it does not display the sub-concept and super-concept words, it can accurately translate keywords.

Some keywords have no sub-concept, so when we search for such words, only the translate words are shown, like Knowledge reasoning.

### 5.4. Examples show

We designed this system to display all keywords information on our database which the user want to know.And this database are big enough. We showed the results of the acquisition of some main keywords like Classification and Computer Vision.

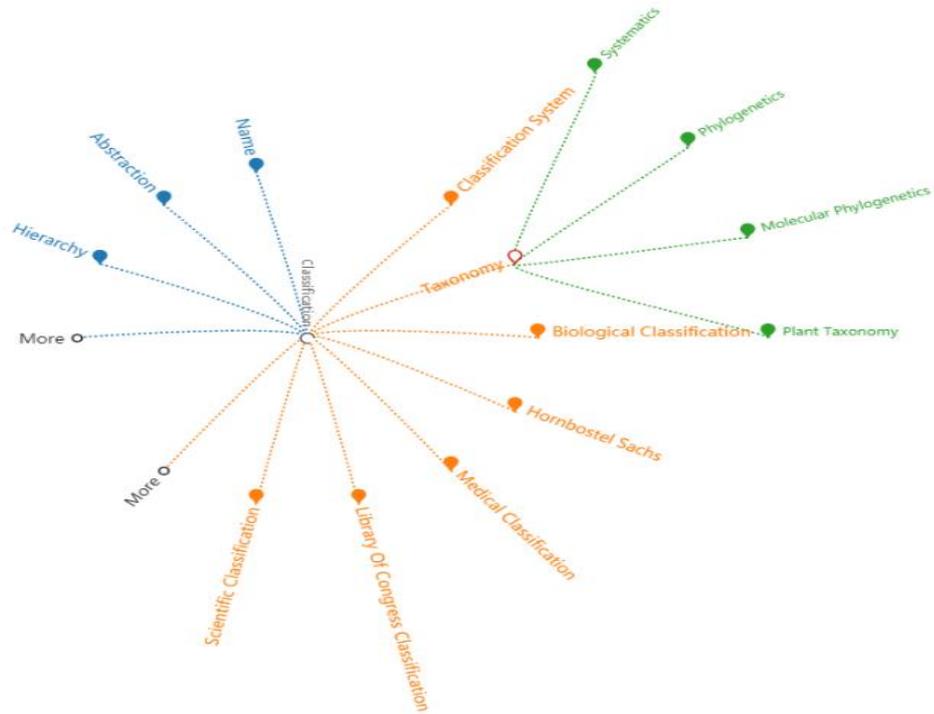

Figure 13. The figure is an example keyword: Classification

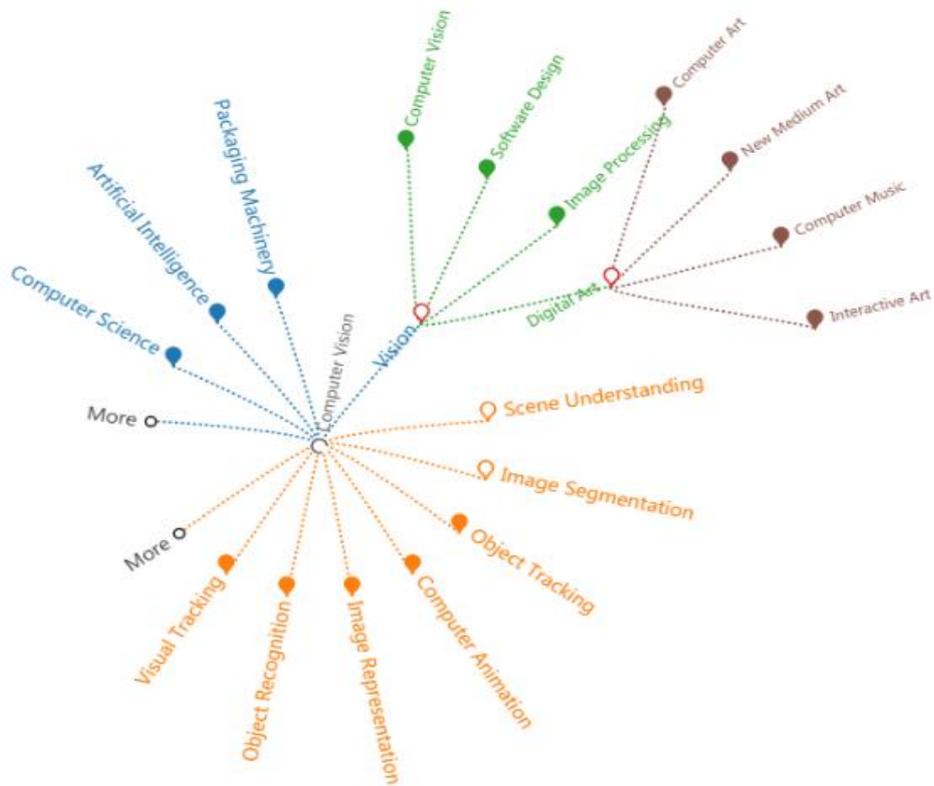

Figure 14. The figure is an example keyword: Computer Vision.

Figure 15. This graph is a system case of visual query.

Figure 16. The figure is a radial tree diagram after expanding more keywords.

Users can change the color of any element in the system by their own needs.

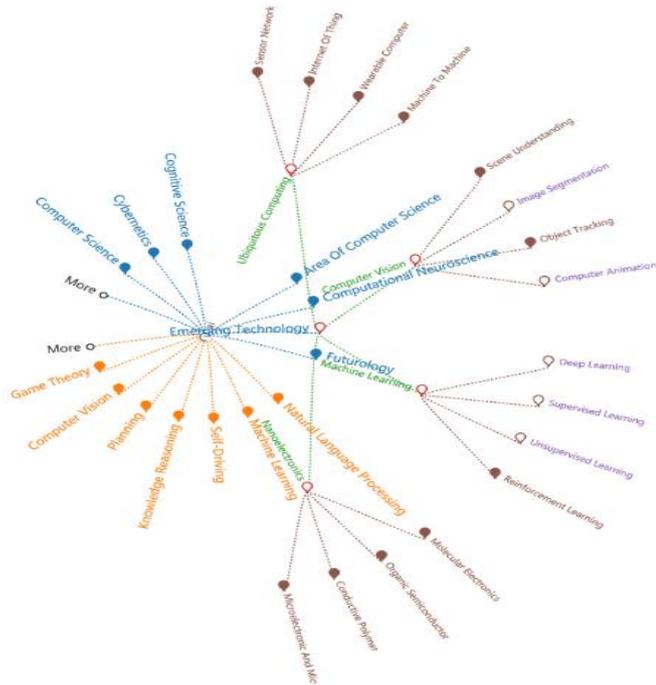

Figure 17. The visual query give us more color change select.

The visualization function not only allows us to click on more sub-nodes we need to expand our visualization, so that the graphics are more hierarchical and the structure is clearer. It contains the differences between various colors and node fills. When more and more nodes become more and more dense, the system will automatically adjust the distance between the connection and the node, making the whole more like a beautiful circle. In the end we formed an overall visual query system for scholar networks. System overview as shown below.

Figure 18. The graph is the full page of our visual query system.

**Author Contributions:** H.Li proposed the idea, made the tool and wrote the paper draft.

**Acknowledgments:** We would like to thank Tsinghua University Aminer group to provide the data and conduct certain survey.

**Conflicts of Interest:** The authors declare no conflict of interest.